# Synthesis, Optoelectronic Properties, and Charge Carrier Dynamics of Colloidal Quasi-two-dimensional $Cs_3Bi_2I_9$ Perovskite Nanosheets†

Sushant Ghimire,*[a] Chris Rehhagen,[a] Saskia Fiedler,[b] Rostyslav Lesyuk,[a] Stefan Lochbrunner,[a] Christian Klinke*[a,c,d]

Non-toxicity and stability make two-dimensional (2D) bismuth halide perovskites better alternatives to lead-based ones for optoelectronic applications and catalysis. In this work, we synthesize sub-micron size colloidal quasi-2D $Cs_3Bi_2I_9$ perovskite nanosheets and study their generation and relaxation of charge carriers. Steady-state absorption spectroscopy reveals an indirect bandgap of 2.07 eV, which is supported by the density functional theory calculated band structure. The nanosheets do not show detectable photoluminescence at room temperature at band-edge excitation which is attributed to the indirect bandgap. However, cathodoluminescence spanning a broad range from 500 nm to 750 nm with an asymmetric and Stokes-shifted spectrum is observed, indicating the phonon- and trap-assisted recombination. We study the ultrafast charge carrier dynamics in $Cs_3Bi_2I_9$ nanosheets using a femtosecond transient absorption spectroscopy. The samples are excited with pump energies higher than their bandgap, and the results are interpreted in terms of hot carrier generation (<1 ps), thermalization with local phonons (~1 ps), and cooling (>30 ps). Further, a relatively slow relaxation of excitons (>3 ns) at the band edge suggests the formation of stable polarons which decay nonradiatively by emitting phonons.

## Introduction

The outstanding improvement in the efficiency of perovskite solar cells from 3.8%[1] to 25.7%[2] in the last thirteen years has demonstrated lead halide perovskites as excellent light-harvesting materials. Besides photovoltaics, lead halide perovskites are also promising for light-emitting diodes (LEDs),[3] photodetectors,[4] lasers,[5] and catalysis.[6] However, the use of lead-based perovskites in optoelectronic devices and catalysis suffers from two limitations. The first one is stability since the three-dimensional (3D) $ABX_3$ [A=Cs, methylammonium (MA, $CH_3NH_3^+$), and formamidinium (FA, $NH_2CH=NH_2^+$); B=Pb; and X=Cl, Br, I] structures are easily disintegrated into their constituent ions by water or other polar solvents and can undergo photodegradation.[7–10] This limitation can be uplifted to some extent by using bulky organic spacer cations such as butylammonium pentylammonium, hexylammonium,[11] octyldiammonium,[12] phenyl ethylammonium,[13] (aminomethyl)piperidinium,[14] and (aminomethyl)pyridinium[15] which form two-dimensional (2D) Ruddlesden-Popper or Dion-Jacobson structure. The inorganic $(PbX_6)^{4-}$ octahedral layers are sandwiched in between and isolated from the surrounding by these hydrophobic spacer layers, rendering better stability against moisture. Another limitation that $APbX_3$ perovskites possess is their toxicity towards the environment and humans.[16] Tin or germanium halide perovskites are not considered a suitable replacement for lead-based ones as they are intrinsically unstable due to their oxidation from +2 to higher (+4) oxidation states.[17,18] Also, $Sn^{2+}$ and $Ge^{2+}$ cannot completely overtake $Pb^{2+}$ in terms of toxicity since the former show similar concerns towards the environment and human health as Pb does.[19] Therefore, a stable and non-toxic alternative is needed for the practical and benign applications of halide perovskites.

Substitution of $Pb^{2+}$ by an isoelectronic $Bi^{3+}$ form a low-toxic $A_3Bi_2X_9$ alternative to $APbX_3$ perovskites. Unlike 3D $APbX_3$, $A_3Bi_2X_9$ are lower dimensional vacancy-ordered perovskites where the additional B-site charge is compensated with ordered vacancies.[20,21] The $(BiX_6)^{3-}$ octahedra are either face-sharing or corner-sharing to form zero-dimensional (0D), one-dimensional (1D), or 2D structures.[20,22] For example, $A_3Bi_2I_9$ consists of isolated $(Bi_2I_9)^{3-}$ dimers of face-sharing octahedra which are arranged hexagonally to form a 0D molecular salt crystal structure. The hexagonal voids are occupied by $Cs^+$ or $MA^+$ cations. Besides, the $Cs_3BiX_6$ (X=Cl, Br, I) polymorphs of $A_3Bi_2X_9$ perovskites in the form of 0D ($Cs_3BiCl_6$ and $Cs_3BiBr_6$) and 3D ($Cs_3BiI_6$) structures are also known.[22] Such a huge structural diversity results in distinctive electronic band features,

[a] Institute of Physics, University of Rostock, Albert-Einstein-Str. 23-24, 18059 Rostock, Germany. E-mail: sushant.ghimire@uni-rostock.de, Christian.klinke@uni-rostock.de
[b] Federal Institute for Materials Research and Testing (BAM), Richard-Willstätter-Str. 11 D-12489 Berlin.
[c] Department of Chemistry, Swansea University, Swansea SA2 8PP, UK.
[d] Department "Life, Light & Matter", University of Rostock, Albert-Einstein-Str. 25, 18059 Rostock, Germany.

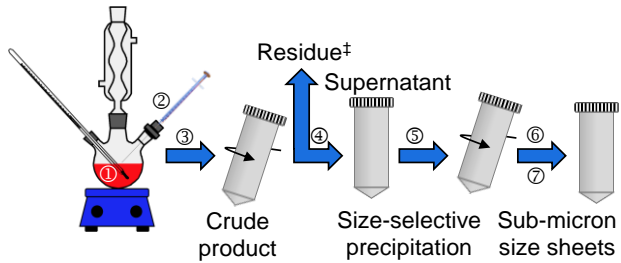

**Fig. 1** Scheme for the synthesis of sub-micron large colloidal quasi-2D Cs$_3$Bi$_2$I$_9$ perovskite nanosheets.

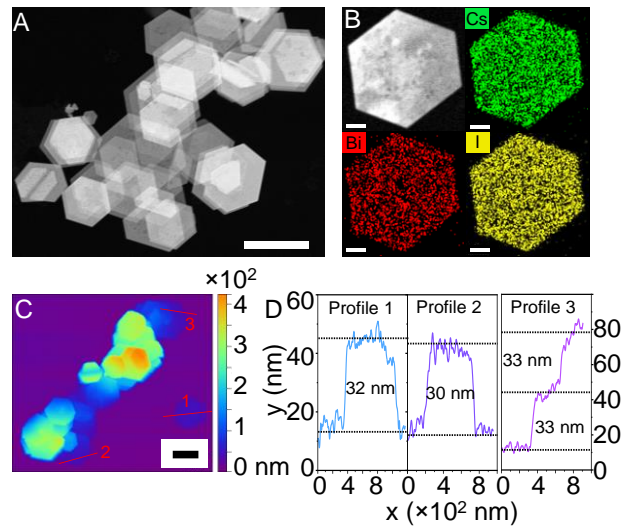

**Fig. 2** Characterization of quasi-2D Cs$_3$Bi$_2$I$_9$ perovskite nanosheets. (A) STEM image of the nanosheets (scale bar 500 nm). (B) Energy-dispersive X-ray (EDX) mapping of a Cs$_3$Bi$_2$I$_9$ nanosheet and the corresponding element maps for Cs, Bi, and I (scale bar 100 nm). (C) Atomic force microscopy (AFM) image of the nanosheets and (D) the corresponding height profiles of the nanosheets that are marked 1, 2, and 3, respectively with red lines in (C).

optoelectronic properties, and defect chemistry in Bi-based perovskites.[23,24]

Among different bismuth-based vacancy-ordered perovskites, A$_3$Bi$_2$I$_9$ (A=Cs, MA) are of particular interest because they are potential light absorbers for photovoltaics and catalysis in the visible spectrum.[25–28] These perovskites have indirect bandgaps.[29] The 0D molecular salt crystal structure on the other hand leads to strong quantum confinement with high carrier effective masses and large exciton binding energy either in their bulk or nanostructured form.[30] Solar cells using A$_3$Bi$_2$X$_9$ bulk film as the light absorber were first reported in 2015 with a power conversion efficiency (PCE) of 1.09% for Cs$_3$Bi$_2$I$_9$, 0.12% for MA$_3$Bi$_2$I$_9$, and 0.003% for MA$_3$Bi$_2$I$_{9-x}$Cl$_x$.[25] More than a half-decade later the current maximum PCE remains at 3.2% which is for Cs$_3$Bi$_2$I$_9$ nanosheet film.[26] The strong quantum confinement of charge carriers in 0D (Bi$_2$I$_9$)$^{3-}$ dimers limits the photoconductivity and carrier diffusion in A$_3$Bi$_2$I$_9$ (A=Cs, MA) perovskites which are detrimental to photovoltaic applications.[31] While the performance of bismuth halide perovskites is still far below that of APbX$_3$ perovskites, there is enough room to utilize these materials in photovoltaics and catalysis by benefitting from their low toxicity, structural diversity, and unique electronic band features. Therefore, synthesis and study of colloidal A$_3$Bi$_2$X$_9$ perovskites are plausible not only for virtually trading toxic APbX$_3$ perovskites with improved performance but also for the fundamental understanding of the optoelectronic processes in them.

In this work, we synthesized sub-micron large colloidal quasi-2D Cs$_3$Bi$_2$I$_9$ perovskite nanosheets, characterized them using scanning transmission electron microscopy (STEM), energy-dispersive X-ray (EDX) spectroscopy, selected area electron diffraction (SAED), and powder X-ray diffraction (XRD). We then examined the electronic band structure of quasi-2D Cs$_3$Bi$_2$I$_9$ perovskite nanosheets theoretically, and studied their optoelectronic properties and charge carrier dynamics using cathodoluminescence (CL) spectroscopy, steady-state UV-vis absorption, and transient absorption spectroscopy. While the CL study reveals broadband emission in these samples, the transient absorption measurements show slow exciton relaxation dynamics (≥3000 ps). Such a slow carrier relaxation suggests strong self-trapping of excitons by the formation of stable polarons.[21] The study of ultrafast carrier dynamics above the band edge of colloidal quasi-2D Cs$_3$Bi$_2$I$_9$ nanosheets reveals hot carrier generation in a sub-picosecond time scale and their cooling within several picoseconds. Tailor et al. have reported such slow hot charge carrier cooling dynamics in A$_3$Bi$_2$I$_9$ (A=Cs, MA, FA) perovskite bulk single crystals and films.[32,33] The delayed exciton recombination in Cs$_3$Bi$_2$I$_9$ can be utilized to harvest charge carriers efficiently for photovoltaics and catalysis.

## Results and Discussion

We used a two-step approach, the hot-injection method followed by selective reprecipitation, to synthesize the sub-micron large colloidal quasi-2D Cs$_3$Bi$_2$I$_9$ perovskite nanosheets. A detailed synthesis scheme is shown in Fig. 1, and the procedure is given in the Experimental Section. In the first step, octylammonium iodide dissolved in γ-butyrolactone (or dimethylformamide) was injected into a mixture of cesium oleate and bismuth oleate in diphenyl ether at 120 °C. The reaction was quenched after 30 s by placing the reaction mixture in an ice bath. The precipitate and the supernatant were separated and collected by centrifugation. In the second step, size-selective reprecipitation was performed by adding isopropyl alcohol dropwise to the supernatant. Here, the precipitate contains nonuniformly size-distributed small and large nanoplates. Characterization of these nonuniform size nanoplates is given in the Supporting Information in Fig. S1. Zhang et al.[34] have reported a nucleation-controlled solution method to grow centimeter size large Cs$_3$Bi$_2$I$_9$ perovskite single crystals where the excess nucleation seeds in the form of sub-millimeter size perovskite crystallites were removed by decantation once they were formed when the reaction mixture was heated at 80 °C for 24 hours. The large single crystals were then grown from the supernatant by heating the system at 80 °C to 95 °C at a rate of 2 °C/day. In our case, we controlled the concentration of nucleation seeds by removing size-distributed nanoplates formed by hot injection in the first step. Further, the supernatant with a critical concentration of nucleation seeds was supersaturated by the addition of isopropyl alcohol. Here, the critical concentration means the optimal concentration of nucleation seeds requires to grow sub-micron large nanosheets. Also, we achieved large colloidal nanosheets in a much shorter time (2-5 min) and at room temperature. We characterized the morphology and elemental composition of the samples by using STEM and EDX spectroscopy (Fig. 2). Fig. 2A shows the STEM image of the quasi-2D Cs$_3$Bi$_2$I$_9$ perovskite nanosheets which reveal a hexagonal shape. The lateral dimensions are in the range of 500 nm. Fig. 2B shows the EDX mapping of the nanosheet which shows uniform distribution of Cs, Bi, and I elements. The corresponding EDX spectrum along with the atomic fraction of the constituent elements are given in the Supporting Information (Fig. S2). Fig. 2C shows the atomic force microscopy (AFM) image of the samples. The corresponding height profiles of the nanosheets are shown in Fig. 2D which reveals an average thickness of 32 nm.

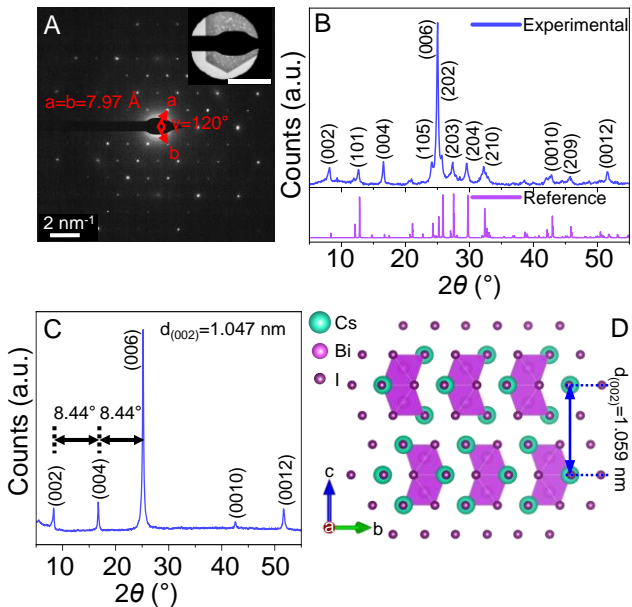

**Fig. 3** Crystal structure of quasi-2D Cs$_3$Bi$_2$I$_9$ perovskite nanosheets. (A) Selected area electron diffraction (SAED) pattern of the nanosheet. The corresponding selected area image is shown in the inset (scale bar 200 nm). The lattice parameters a=b=7.97 Å are obtained as distances measured from the center of the SAED pattern along a and b directions that extend at an angle of γ=120° for the hexagonal unit cell. (B) Powder XRD pattern of the nanosheets. The diffraction peaks are labelled with the Miller indices by comparing them with the reference pattern PDF card 01-073-0707 of a hexagonal Cs$_3$Bi$_2$I$_9$ crystal structure. (C) XRD pattern of the nanosheet film showing periodicity of 2θ=8.44° and d-spacing of 1.047 nm along (002) plane. (D) A 1×3×1 supercell crystal structure of Cs$_3$Bi$_2$I$_9$ perovskites obtained from the PDF card 01-073-0707 showing a d-spacing of 1.059 nm along (002) plane.

We studied the crystal structure of the quasi-2D Cs$_3$Bi$_2$I$_9$ nanosheets by SAED and powder XRD, which are shown in Fig. 3. The SAED reveals sharp and hexagonally-ordered lattice diffraction spots, indicating high crystalline quality (Fig. 3A). Further, the XRD pattern of quasi-2D Cs$_3$Bi$_2$I$_9$ nanosheet powder (Fig. 3B) reveals a hexagonal crystal structure with a P 63/m m c space group which is in agreement with Powder Diffraction File (PDF) card number 01-073-0707 and Cs$_3$Bi$_2$I$_9$ single crystal reported by Zhang et al.[34] However, the diffraction peaks are quite shifted (~0.18°) towards lower 2θ angle, indicating the existence of lattice strain in our colloidal nanosheets. Here, a prominent (006) diffraction peak indicates that the crystal has preferential growth along the c-axis. Fig. 2C is the XRD pattern of the quasi-2D Cs$_3$Bi$_2$I$_9$ nanosheets in a film which shows diffraction peaks with (002), (004), (006), (0010), and (0012) Miller indices.[34] The diffraction peaks corresponding to (002), (004), and (006) are equally spaced with a periodicity of 8.44° 2θ angle, indicating the preferential orientation of the nanosheets in the film. From the periodicity, a d-spacing is calculated as 1.047 nm which is marginally smaller than the reference data [d$_{(002)}$=1.059 nm] obtained from the PDF card 01-073-0707. Fig. 3D shows the 1×3×1 supercell crystal structure of Cs$_3$Bi$_2$I$_9$ perovskites. These perovskites have isolated 0D [Bi$_2$I$_9$]$^{3-}$ dimers formed by face-sharing of [BiI$_6$]$^{3-}$ octahedra which are surrounded by the (Cs$^+$) A-site cations. The [Bi$_2$I$_9$]$^{3-}$ dimers are in a layered arrangement along the ab (00l) plane.[34] We determined the lattice parameters to be a=b=7.97 Å and γ=120° from the SAED pattern. Further, taking d$_{(002)}$=10.47 Å and using eqn (1),

$$d_{(hkl)} = \frac{1}{\sqrt{\left(\frac{4}{3a^2}\right)(h^2+k^2+hk)+\left(\frac{l^2}{c^2}\right)}} \qquad (1),$$

for a hexagonal crystal structure, we calculated c=20.94 Å. These lattice parameters are moderately smaller than the values reported for 3D single crystals of Cs$_3$Bi$_2$I$_9$,[34] which further supports the development of lattice strain in our sample. We anticipate that the lower dimensional structure and the involvement of shorter-chain octylammonium ligands induce lattice strain in our samples. Such lattice strain results in octahedral distortion in halide perovskites which influences their optoelectronic properties.[35] We recently reported octahedral distortion in 2D octylammonium tin iodide perovskites, resulting in structural reconstruction, formation of self-trapped excitons, and broadband emission.[36]

We studied the optical and electronic properties of quasi-2D Cs$_3$Bi$_2$I$_9$ nanosheets using steady-state UV-vis absorption spectroscopy and CL microspectroscopy, which are shown in Fig. 4. The absorption spectrum in Fig. 4A shows a broad absorption onset and a sharp excitonic peak at 495 nm. While the sharp absorption peak is the result of strong confinement of excitons in the isolated 0D [Bi$_2$I$_9$]$^{3-}$ dimers, the broad absorption onset is the result of an indirect bandgap.[29] We also do not rule out the presence of surface defects in these nanosheets which contribute to the tail in the absorption spectrum.[37] The direct and indirect bandgap values are obtained as 2.35 eV and 2.07 eV, respectively from the Kubelka-Munk plot, as shown in Fig. 4B. The indirect bandgap nature of Cs$_3$Bi$_2$I$_9$ perovskites results in strong electron-phonon coupling.[21] As a result, our samples do not show detectable photoluminescence at room temperature at band-edge excitation. Previous studies have reported luminescent Cs$_3$Bi$_2$I$_9$ perovskites at room temperature attributing to self-trapped excitons and defect emission,[21,38] while others show photoluminescence only at cryogenic temperatures.[39,40] On the contrary, some reports also show that the photoluminescence in Cs$_3$Bi$_2$X$_9$ (X=Cl, Br, I) at room temperature originates not from the perovskite structure itself but the precursor salt BiX$_3$.[41] We recorded the excitation spectrum of the sample, as shown in the Supporting Information in Fig. S3, which is not identical to the absorption spectrum and shows no prominent excitation band at the bandgap. Instead, an excitation band in the deep UV region (<250 nm) is observed. This further clarifies why our samples do not emit when photoexcited at the band edge.

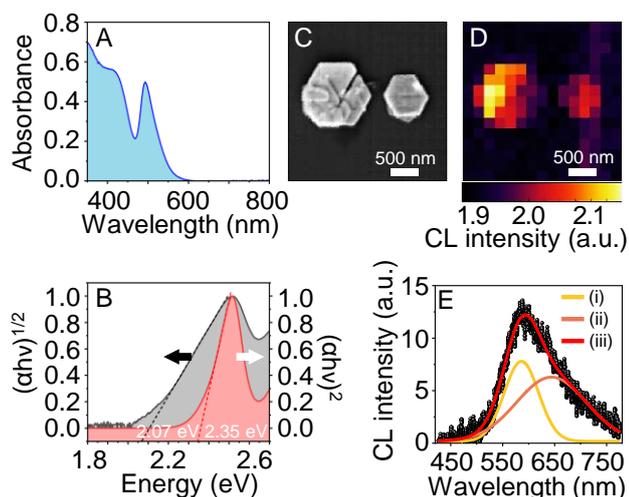

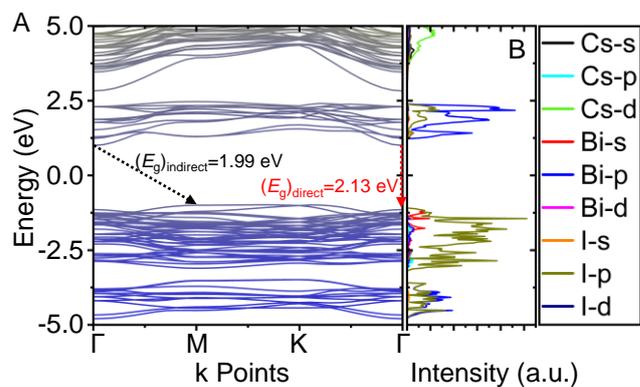

**Fig. 4** Optoelectronic properties of quasi-2D Cs$_3$Bi$_2$I$_9$ perovskite nanosheets. (A) The UV-visible absorption spectrum of the nanosheets as colloidal solution. (B) A Kubelka-Munk plot from the absorption spectrum showing a direct (2.35 eV) and an indirect (2.07 eV) bandgap values. (C-E) Cathodoluminescence (CL) study of Cs$_3$Bi$_2$I$_9$ nanosheets showing (C) SEM image, (D) the corresponding 2D spatial maps of CL intensity and (E) a representative CL spectrum. The CL spectrum is fitted for multiple peaks with Gaussian function and the fitted spectra are also shown.

**Fig. 5** Electronic band structure of quasi-2D Cs$_3$Bi$_2$I$_9$ perovskite nanosheets. (A) Electronic band structure using DFT-GGA calculation and self-consistent SOC. (B) The corresponding partial density of states (DOS).

To further clarify the luminescence properties of quasi-2D Cs$_3$Bi$_2$I$_9$ perovskite nanosheets, we performed SEM-CL microspectroscopy at 7.0 keV electron beam acceleration voltage and 500 pA electron beam current. Fig. 4C and D show the secondary electron (SE) image and the corresponding CL intensity map, respectively. The non-uniform distribution of CL intensities in Fig. 4D is most likely due to the stacking of multiple nanosheets and differences in their surface texture. We also do not rule out the crystallographic defect-related CL intensity distribution.[42–44] Nevertheless, the defect-related CL was not resolved in the intensity map which is also supported by the similar panchromatic CL maps of the samples without and with (500/50 nm and 600/50 nm) bandpass filters that are provided in the Supporting Information in Fig. S4. Fig. 4E shows a typical CL spectrum of a single quasi-2D Cs$_3$Bi$_2$I$_9$ nanosheet which reveals a broad and asymmetric shape. The broadening of CL spectra in the case of lead-based 2D halide perovskites is correlated to the electron beam-induced defect formation and the thermal effects induced by the population of higher vibrational modes due to high-energy electron excitation.[45] The CL spectrum in our case is deconvoluted by Gaussian fitting into two spectra, one of which is relatively narrow (full width at half maximum, FWHM=78 nm) and centered at 585 nm while the other is broad (FWHM=158 nm) and centered around 644 nm. The CL at 585 nm (2.11 eV) is close to the bandgap of the material and is considered to be originated from the band-edge recombination of excitons. On the other hand, CL at 644 nm (1.92 eV) is anticipated from a defect- or trap-assisted recombination which agrees well with the observations by Gosh et al.[31] They have reported a broad and asymmetric CL spectrum in Cs$_3$Bi$_2$I$_9$ thin film which was resolved into two peaks, one at 2.5 eV which they assigned to the direct exciton emission, and another at 1.91 eV which was assigned to the emission from self-trapped excitons by extrinsic defects.

We calculated the electronic band structure and partial density of states (DOS) of quasi-2D Cs$_3$Bi$_2$I$_9$ nanosheets using the experimental crystal structure and a Density Functional Theory (DFT)[46,47] with the Generalized Gradient Approximation (GGA)[48] functional and self-consistent spin-orbital coupling (SOC). The band structure shown in Fig. 5A is consistent with the earlier report[49] and well supports the indirect bandgap nature of these materials. Here, a minimum of the conduction band occurs at Γ point and the maximum of the valence band occurs at the M point of the Brillouin zone. The theoretical indirect bandgap (1.99 eV) is in close agreement with the experimental value. Partial DOS of quasi-2D Cs$_3$Bi$_2$I$_9$ nanosheets is shown in Fig. 5B which reveals that the valence band maximum is mainly formed by I 5p-Bi 6s hybrid orbitals and the conduction band minimum is formed by Bi 6p-I 5p hybrid orbitals. Nevertheless, Cs-based orbitals do not contribute to the band edge and only form the higher conduction band. Further, in those hybrid orbitals at the band edge, the I 5p and Bi 6p orbitals have a greater contribution to the valence band maximum and conduction band minimum, respectively. This picture of the band structure in quasi-2D Cs$_3$Bi$_2$I$_9$ greatly supports the 0D molecular crystal structure of these materials since the band edge is formed by the hybridization of Bi and I orbitals, and therefore the charge carriers generated must be confined within these hybrid orbitals. In other words, the excitons in Cs$_3$Bi$_2$I$_9$ perovskites are generated and confined within the isolated 0D [Bi$_2$I$_9$]$^{3-}$ dimers.

We studied the charge carrier dynamics in quasi-2D Cs$_3$Bi$_2$I$_9$ nanosheets at different excitation energies and intensities using a femtosecond transient absorption spectroscopy (Fig. 6 and Supporting Information Fig. S5). The transient absorption dynamics of the samples excited at particular pump energy but with different excitation intensities were comparable (Supporting Information Fig. S5) for the measurement time window (~2 ns) and the excitation intensity range (30-200 µW) we used. Therefore, our discussion is focused on the pump energy-dependent transient absorption studies of the quasi-2D Cs$_3$Bi$_2$I$_9$ nanosheet samples. Fig. 6A and B show the transient absorption spectra of the samples excited with the pump energy of 3.10 eV (400 nm) and 2.48 eV (500 nm) with comparable excitation intensities, respectively. These transient absorption spectra show three distinct features: a negative signal (photobleach) at 2.5 eV (495 nm) and two positive bands on either side. The photobleach at 2.5 eV matches with the steady-state absorption peak (495 nm) and is considered a ground-state photobleach (GSPB). On the other hand, a broad positive signal at energy lower than GSPB (>520 nm) is the excited state absorption (ESA) and is assigned as photoinduced absorption (PA) 1. Also a second photobleach is observed at 2.9 eV (420 nm), which corresponds to the second excitonic transition (a hump) present in the steady-state absorption spectrum in Fig. 4A.[49,50] We fitted these transient absorption spectra globally using a stretched exponential function given in eqn (2),

$$n(t) = n_0\, exp(-\sqrt{t/\tau})  \qquad (2).$$

**Table 1** Parameters obtained from the global fitting of transient absorption spectra of quasi-2D Cs$_3$Bi$_2$I$_9$ perovskite nanosheets using a stretched exponential function given in eqn 2.

| Pump energy | A1 | τ$_1$ | A2 | τ$_2$ | A3 | τ$_3$ |
|---|---|---|---|---|---|---|
| 3.10 eV | 1 | 1.40 ps | 1 | 73 ps | 1 | 3398 ps |
| 2.48 eV | 1 | 1.54 ps | 1 | 29 ps | 1 | 3241 ps |

The results are shown in Table 1 and Fig. 6C-F. The same amplitudes for all three-lifetime components indicate that the GSPB and ESA are dominant for the entire observable time window. The long carrier lifetime (>3 ns) is assigned to the relaxation of excitons at the indirect bandgap, while the short (~1.5 ps) and intermediate (~30-70 ps) lifetime components comprise the intraband cooling of hot charge carriers which are generated via excitation of the nanosheets above their bandgap. The very short carrier lifetimes are comparable for low (2.48 eV) and high (3.1 eV) pump energy. On the other hand, the intermediate and the long carrier lifetime components vary for these two pump energies (Table 1). This is well reflected in Fig. 6C where the GSPB recovery for the 3.10 eV pump starts much later when compared to that for the 2.48 eV pump. Nevertheless, the overlapping of GSPB with broad PA1 might also influence the intensity of the photobleach, especially in the case of higher pump energy (3.10 eV). The PA1 for 3.10 eV pump energy is much broader compared to the one for 2.48 eV which is attributed to the distribution of hot charge carriers over a larger number of higher energy states above and below the conduction band minima and the valence band maxima, respectively. This is obvious since more energy states are accessible to the hot carriers when excited with higher pump energy. Unlike PA 1, the positive signal (PA2) at 470 nm wavelength (i.e. energy higher to the GSPB) is much stronger and spectrally well-defined. Its position corresponds to a dip present between the first and the second excitonic transitions in the steady-state absorption spectrum. The origin of PA2 is different from the origin of PA1, as revealed by the bleach reconstructed from the steady-state absorption spectrum shown in the evolution-associated spectra in Fig. 6E and F and a second derivative of the steady-state absorption spectrum shown in the Supporting Information in Fig. S6. In both cases, only one positive band at 470 nm is prominently observed along with a negative photobleaching signal. This suggests that PA2 is not associated with the intraband transition of the charge carriers created by the pump pulse, while PA1 is the ESA of pump pulse-created charge carriers. Rosi et al. have discussed the origin of such photoinduced absorption bands in transient absorption spectra of strongly confined quantum dots concerning light-induced activation of parity-forbidden exciton transition enabled by the formation of symmetry-breaking polarons after the excitation of excitons taking CsPbBr$_3$ perovskites as an example.[50] They argue that the intraband transition occurs at lower energy than the exciton transition and therefore does not justify the appearance of PA in the 2.5–3 eV range. Such a photoinduced forbidden transition of excitons from the valence band to the conduction band might be the origin of PA2 in our case too. Nevertheless, a clear picture in this regard is still elusive and requires further investigations.

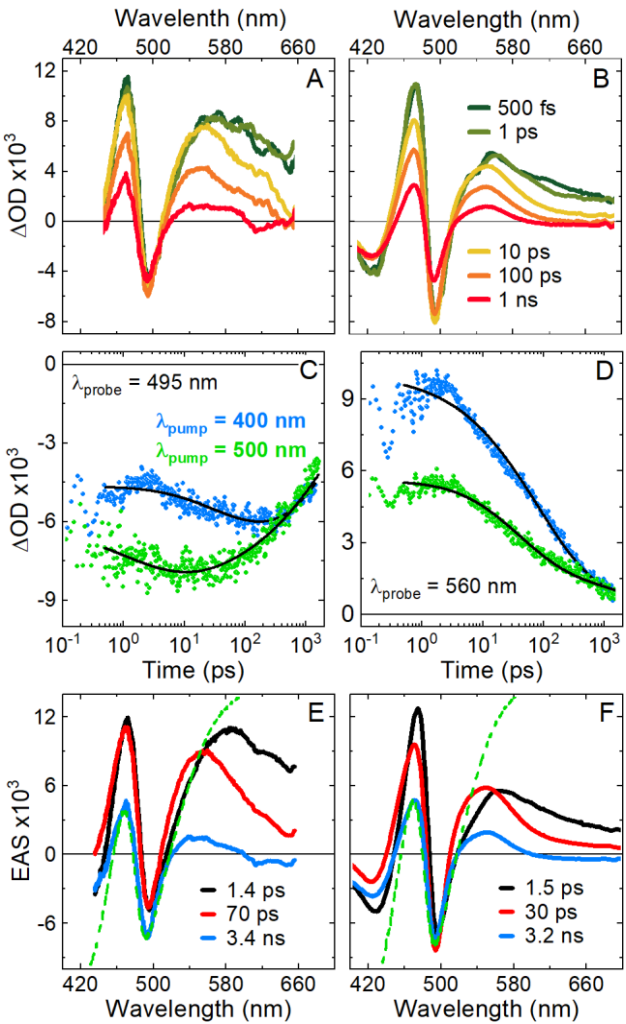

**Fig. 6** Ultrafast transient absorption studies of quasi-2D Cs$_3$Bi$_2$I$_9$ perovskite nanosheets. (A, B) Transient absorption spectra at pump energies (A) 3.1 eV (400 nm) and (B) 2.48 eV (500 nm). (C, D) Kinetic traces at different pump energies for (C) ground state bleach (λ$_{probe}$=495 nm) (D) excited-state absorption (λ$_{probe}$=560 nm). Fits are shown in black. (E, F) Evolution associated spectra (EAS) at pump energies (E) 3.1 eV and (F) 2.48 eV obtained from global fitting of the transient absorption spectra with three stretched exponential fit components as shown in the insets. The dotted lines show the bleach reconstructed from the absorption normalized to the minimum of the signal.

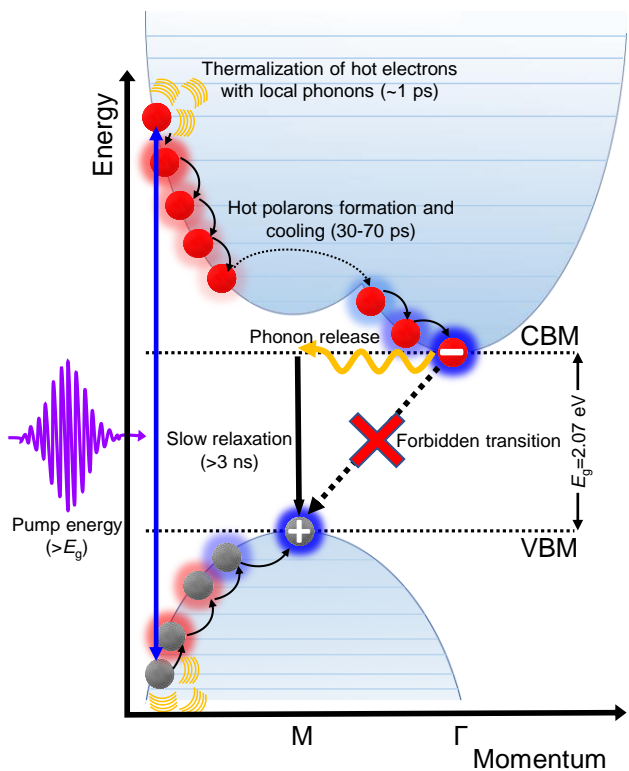

**Fig. 7** Proposed mechanism of charge carrier relaxation in quasi-2D Cs$_3$Bi$_2$I$_9$ perovskite nanosheets.

A proposed mechanism of charge carrier relaxation in quasi-2D Cs$_3$Bi$_2$I$_9$ nanosheets is shown schematically in Fig. 7. When the nanosheets are excited with photon energy larger than their bandgap, highly energetic charge carriers are generated deep into the conduction and valence bands, which are called hot charge carriers. The generation of hot charge carriers happens within a sub-picosecond time scale. These hot electrons and holes then thermalize with the local lattice phonons which envelop these charge carriers and form hot polarons. In our case, the short dynamics of 1.5 ps can be assigned to this thermalization process. Afterward, the hot polarons equilibrate with the lattice by cooling down to the band edge and turning into cold polarons. In our case, this happened within 30-70 ps depending upon the energy of excitation. The longer time (>70 ps) in the case of 3.10 eV pump energy when compared to the 2.48 pump (30 ps) can be attributed to higher hot carrier temperature in the case of the former which takes a relatively long time to cool down. Finally, once the cold polarons reach the conduction band minima and the valence band maxima at the indirect bandgap, they relax to the ground state by emitting phonons. In our case, this relaxation is delayed and happens in >3 ns time scale. Frost et al. have discussed such a formation and cooling of hot polarons in direct bandgap MAPbI$_3$ perovskites. They have reported a typical time scale of 1-100 fs for hot carrier formation, 1-10 ps for their thermalization and hot polaron formation, and 10-100 ps for hot polaron cooling.[51] Further, Wong et al. have reported a very long lifetime of >4 ns for the polaron population decay in 3D MAPbI$_3$ perovskites.[52] The electron-phonon coupling and formation of polarons not only affect the charge carrier dynamics in Cs$_3$Bi$_2$I$_9$ perovskites but also influence their luminescence properties. Although we could not detect photoluminescence in our samples, the observation of broad and asymmetric CL from them suggests electron-phonon coupling and the self-trapping of polarons. McCall et al. have demonstrated strong electron-phonon interactions in A$_3$B$_2$I$_9$ (A=Cs, Rb and B=Bi, Sb) perovskites which induce the formation of small polarons.[21] These polarons are self-trapped within the lattice resulting in broadband emission. These reported trends of hot carrier generation, cooling, and relaxation greatly support our current findings and proposed mechanism except that in their case, the final relaxation of charge carriers emits photons while in our case, it is nonradiative and emits phonons.

## Experimental

### Materials

All reagents were commercially purchased and used without further purification: cesium acetate, bismuth acetate, octylammonium iodide, oleic acid, diphenyl ether, toluene, γ-butyrolactone, and dimethyl formamide (DMF).

### Methods

#### Synthesis of quasi-2D Cs$_3$Bi$_2$I$_9$ nanosheets

In a typical hot-injection method, cesium acetate (0.1 mmol, 20 mg), bismuth acetate (0.1 mmol, 40 mg), and oleic acid (3.16 mmol, 1 mL) were dissolved and dried in diphenyl ether (10 mL) in a three-neck flask under stirring and high vacuum at 80 °C for 1.5 h. Afterward, under argon flush, the temperature of the reaction mixture was increased to 120 °C and n-octylammonium iodide (1.6 mmol, 411 mg) dissolved in γ-butyrolactone (300 μL) was injected. The reaction was quenched by placing the reaction mixture in an ice bath after 30 s. After quenching, it was centrifuged, and the orange-colored supernatant and precipitate were separated. The precipitate was washed three times by repeatedly dispersing it in toluene and centrifuging at 10000 rpm for 10 min. It was finally collected for further characterization. The supernatant was subjected to a size-selective reprecipitation of the sub-micron large quasi-2D $Cs_3Bi_2I_9$ nanosheets by adding 3-5 mL of isopropyl alcohol dropwise into it and letting it stand undisturbed for about 15 min. Purification of the selectively precipitated large nanosheets was carried out by their repeated centrifugation and washing with toluene as shown in Fig. 1.

**Characterization of quasi-2D $Cs_3Bi_2I_9$ nanosheets**

We characterized the morphology, elemental composition, thickness, and crystal structure of $Cs_3Bi_2I_9$ samples using scanning transmission electron microscopy (STEM, ThermoFisher Talos L120C) equipped with an energy-dispersive X-ray (EDX) detector, atomic force microscopy (AFM), and X-ray diffraction (XRD, Aeris X-ray diffractometer, Malvern Panalytical). TEM samples were prepared on carbon-coated copper grids and observed at an acceleration voltage of 200 kV. Selected area electron diffraction (SAED) was also performed on the individual nanosheet in TEM. For AFM, samples were prepared by dropping a dilute solution of the nanosheet on a silicon wafer which was then mounted on a stainless steel holder through a double-sided adhesive tape. Powder XRD was performed on dried powdered nanosheet samples as well as the corresponding film deposited on a low-background silica disc. The wavelength of the Cu-kα1 X-ray radiation was 1.5406 Å.

**Measurements of optoelectronic properties**

**Ultraviolet-visible (UV-Vis) absorption and photoluminescence excitation (PLE) spectroscopy**

UV-Vis absorption spectra were recorded using a Lambda 1050+ UV-Vis-NIR spectrometer from Perkin Elmer. PLE spectra were recorded on a spectrofluorometer FS5 from Edinburg Instrument. For the PLE measurement, a virtual emission wavelength at 600 nm (which corresponds to the absorption onset) was set, and the excitation scan was performed in a range of 200 nm to 550 nm with an emission and excitation slit width of 5 and 3 nm, respectively. Both the UV-vis and the PLE spectra were measured in a 1 cm x 1 cm x 1 cm quartz cuvette filled with the colloidal solution of quasi-2D $Cs_3Bi_2I_9$ nanosheets in toluene.

**Cathodoluminescence (CL) spectroscopy**

Cathodoluminescence measurements were performed in a Zeiss Merlin scanning electron microscope equipped with a DELMIC Spark CL spectrograph. All spectra were collected at an electron beam energy of 7 keV and a current of 500 pA. To counteract the bleaching of the sample, the electron beam was moved during the 30 s exposure time.

**Band structure calculation**

The density functional theory (DFT) package, ABINIT package,[46,47] a joint project of the Université Catholique de Louvain, Corning Incorporated, and other contributors (URL http://www.abinit.org) was used for band structure and density of states calculations. As exchange functional, Generalized Gradient Approximation (GGA)[48] and Hartwigsen-Goedecker-Hutter pseudopotentials[53] were used. The lattice parameters were fixed to the experimental values.

**Femtosecond transient absorption spectroscopy**

For femtosecond transient absorption spectroscopy two set-ups were used depending on the desired excitation wavelength. Both set-ups have in common, that the output of a 1 kHz regenerative Ti: sapphire amplifier laser system is split for the generation of a probe and a pump beam. As a probe beam, a supercontinuum white light ranging from 350nm to 750 nm was generated in a $CaF_2$ crystal. The pump beam was obtained by either frequency doubling or pumping a noncollinear optical parametric amplifier. In the latter case, the pulse was temporally compressed by a fused-silica prism compressor. The intensity was adapted by two wire grid polarizers. The polarization of the pump and the probe were set at a magical angle orientation relative to each other by a half-wave plate. Additional data at parallel and vertical polarization is acquired. Both beams were focused into a 1 mm thick quartz cuvette filled with the sample with the optical density of 0.13 (at 400nm) and 0.11 (at 500nm). The cuvette was rotated with a frequency of 100 per min to avoid photodegradation. The transmitted probe beam was spectrally dispersed by a prism and detected by a photodiode array detector. The pump light was chopped with half of the repetition rate of the probe beam allowing us to calculate transient absorption by comparing the absorption with and without excitation by the pump beam.

For excitation at 500 nm (2.48 eV), a NOPA was pumped by the laser system output (CPA 2001, Clark MXR, Inc.). Its output pulses were compressed by the prism compressor to 40 fs (full width at half maximum, FWHM) pulse duration. The pump and probe beams were focused on overlapping spots of 180 μm and 75 μm (FWHM values). For excitation at 400 nm (3.10 eV), the laser system fundamental at 800 nm center wavelength was frequency doubled. The focus sizes of the pump and probe beams were 110 μm and 20 μm (FWHM values). The pump was chopped at a quarter and the probe at half of the repetition rate of the laser system (Spitfire Ace PA, Spectra Physics). Using the resulting four signals for multishot-referencing increases the signal-to-noise ratio of acquired data.

The time resolution was 100 fs for both of the set-ups.

## Conclusions

In summary, a critical concentration of nucleation seeds is needed for the growth of sub-micron large colloidal quasi-2D $Cs_3Bi_2I_9$ perovskite nanosheets which was achieved in our case by removing smaller and size-distributed nanoplates obtained from the hot injection method. Subsequent supersaturation resulting from the addition of isopropyl alcohol precipitated large hexagonal nanosheets of $Cs_3Bi_2I_9$. These morphologically 2D $Cs_3Bi_2I_9$ perovskites have hexagonal crystal structures comprising molecularly 0D isolated dimers of $BiX_6$ octahedra in which excitons are confined. The indirect bandgap of the nanosheets results in no detectable PL when excited at the band edge. However, phonon-assisted recombination of excitons results in a broad and asymmetric CL spectrum. Ultrafast charge carrier dynamics studied via a femtosecond transient absorption spectroscopy reveals the generation and cooling of hot charge carriers in sub-picosecond to tens of picosecond time scale and the relaxation of excitons to the ground state in nanosecond time scale. Such a prolonged time for exciton recombination can be viewed in terms of the generation and cooling of hot polarons and their decay via phonon emission. This study provides detailed insights into the fundamental aspect of optoelectronic properties and charge-carrier dynamics in lead-free bismuth halide perovskites, where the slow exciton dynamics can be utilized to harvest the charge carriers for catalysis.

## Author Contributions

S.G. and C.K. perceived the idea, and S.G. wrote the manuscript. All authors participated in performing experiments, analyzing data, discussing the results, and commenting on the manuscript.

## Conflicts of interest

There are no conflicts of interest to declare.

## Acknowledgments

S.G. acknowledges Alexander von Humboldt-Stiftung/Foundation for the postdoctoral research fellowship. C.K. acknowledges the European Regional Development Fund of the European Union for funding the X-ray diffractometer (GHS-20-0036/P000379642) and the Deutsche Forschungsgemeinschaft for funding an electron microscope ThermoFisher Talos L120C (INST 264/188-1 FUGG) and for supporting the collaborative research center LiMatI (SFB 1477). We acknowledge Prof. Dr. S. Speller of the University of Rostock for providing/supporting us with the Atomic Force Microscope, and Dr. Klaus Schwarzburg of Institute for Solar Fuels, Helmholtz-Zentrum Berlin für Materialien und Energie GmbH, Hahn-Meitner-Platz 1, 14109 Berlin, Germany for supporting us with CL measurements.

## Notes and references

1 A. Kojima, K. Teshima, Y. Shirai and T. Miyasaka, *J. Am. Chem. Soc.*, 2009, **131**, 6050–6051.
2 Best Research-Cell Efficiency Chart, https://www.nrel.gov/pv/cell-efficiency.html, (accessed October 12, 2022).
3 H. Chen, L. Fan, R. Zhang, C. Bao, H. Zhao, W. Xiang, W. Liu, G. Niu, R. Guo, L. Zhang and L. Wang, *Adv. Opt. Mater.*, 2020, **8**, 1901390.
4 J. Feng, C. Gong, H. Gao, W. Wen, Y. Gong, X. Jiang, B. Zhang, Y. Wu, Y. Wu, H. Fu, L. Jiang and X. Zhang, *Nat. Electron.*, 2018, **1**, 404–410.
5 Q. Zhang, Q. Shang, R. Su, T. T. H. Do and Q. Xiong, *Nano Lett.*, 2021, **21**, 1903–1914.
6 H. Huang, B. Pradhan, J. Hofkens, M. B. J. Roeffaers and J. A. Steele, *ACS Energy Lett.*, 2020, **5**, 1107–1123.
7 J. M. Frost, K. T. Butler, F. Brivio, C. H. Hendon, M. van Schilfgaarde and A. Walsh, *Nano Lett.*, 2014, **14**, 2584–2590.
8 L. Chouhan, S. Ghimire and V. Biju, *Angew. Chem. Int. Ed.*, 2019, **58**, 4875–4879.
9 Y. A. Darmawan, M. Yamauchi and S. Masuo, *J. Phys. Chem. C*, 2020, **124**, 18770–18776.
10 P. Toloueinia, H. Khassaf, A. Shirazi Amin, Z. M. Tobin, S. P. Alpay and S. L. Suib, *ACS Appl. Energy Mater.*, 2020, **3**, 8240–8248.
11 I. Spanopoulos, I. Hadar, W. Ke, Q. Tu, M. Chen, H. Tsai, Y. He, G. Shekhawat, V. P. Dravid, M. R. Wasielewski, A. D. Mohite, C. C. Stoumpos and M. G. Kanatzidis, *J. Am. Chem. Soc.*, 2019, **141**, 5518–5534.
12 X. Li, J. Hoffman, W. Ke, M. Chen, H. Tsai, W. Nie, A. D. Mohite, M. Kepenekian, C. Katan, J. Even, M. R. Wasielewski, C. C. Stoumpos and M. G. Kanatzidis, *J. Am. Chem. Soc.*, 2018, **140**, 12226–12238.
13 I. C. Smith, E. T. Hoke, D. Solis-Ibarra, M. D. McGehee and H. I. Karunadasa, *Angew. Chem. Int. Ed.*, 2014, **53**, 11232–11235.
14 L. Mao, W. Ke, L. Pedesseau, Y. Wu, C. Katan, J. Even, M. R. Wasielewski, C. C. Stoumpos and M. G. Kanatzidis, *J. Am. Chem. Soc.*, 2018, **140**, 3775–3783.
15 X. Li, W. Ke, B. Traoré, P. Guo, I. Hadar, M. Kepenekian, J. Even, C. Katan, C. C. Stoumpos, R. D. Schaller and M. G. Kanatzidis, *J. Am. Chem. Soc.*, 2019, **141**, 12880–12890.
16 I. R. Benmessaoud, A.-L. Mahul-Mellier, E. Horváth, B. Maco, M. Spina, H. A. Lashuel and L. Forró, *Toxicol. Res.*, 2016, **5**, 407–419.
17 T. Krishnamoorthy, H. Ding, C. Yan, W. Lin Leong, T. Baikie, Z. Zhang, M. Sherburne, S. Li, M. Asta, N. Mathews and S. G. Mhaisalkar, *J. Mater. Chem. A*, 2015, **3**, 23829–23832.
18 D. Ricciarelli, D. Meggiolaro, F. Ambrosio and F. De Angelis, *ACS Energy Lett.*, 2020, **5**, 2787–2795.
19 A. Babayigit, D. Duy Thanh, A. Ethirajan, J. Manca, M. Muller, H.-G. Boyen and B. Conings, *Sci. Rep.*, 2016, **6**, 18721.
20 K. M. McCall, C. C. Stoumpos, O. Y. Kontsevoi, G. C. B. Alexander, B. W. Wessels and M. G. Kanatzidis, *Chem. Mater.*, 2019, **31**, 2644–2650.
21 K. M. McCall, C. C. Stoumpos, S. S. Kostina, M. G. Kanatzidis and B. W. Wessels, *Chem. Mater.*, 2017, **29**, 4129–4145.
22 S. E. Creutz, H. Liu, M. E. Kaiser, X. Li and D. R. Gamelin, *Chem. Mater.*, 2019, **31**, 4685–4697.
23 Z. Jin, Z. Zhang, J. Xiu, H. Song, T. Gatti and Z. He, *J. Mater. Chem. A*, 2020, **8**, 16166–16188.
24 J. K. Pious, C. Muthu and C. Vijayakumar, *Acc. Chem. Res.*, 2022, **55**, 275–285.
25 B.-W. Park, B. Philippe, X. Zhang, H. Rensmo, G. Boschloo and E. M. J. Johansson, *Adv. Mater.*, 2015, **27**, 6806–6813.
26 F. Bai, Y. Hu, Y. Hu, T. Qiu, X. Miao and S. Zhang, *Sol. Energy Mater. Sol. Cells*, 2018, **184**, 15–21.
27 S. S. Bhosale, A. K. Kharade, E. Jokar, A. Fathi, S. Chang and E. W.-G. Diau, *J. Am. Chem. Soc.*, 2019, **141**, 20434–20442.
28 Z.-L. Liu, R.-R. Liu, Y.-F. Mu, Y.-X. Feng, G.-X. Dong, M. Zhang and T.-B. Lu, *Solar RRL*, 2021, **5**, 2000691.
29 Y. Zhang, J. Yin, M. R. Parida, G. H. Ahmed, J. Pan, O. M. Bakr, J.-L. Brédas and O. F. Mohammed, *J. Phys. Chem. Lett.*, 2017, **8**, 3173–3177.
30 M. Pazoki, M. B. Johansson, H. Zhu, P. Broqvist, T. Edvinsson, G. Boschloo and E. M. J. Johansson, *J. Phys. Chem. C*, 2016, **120**, 29039–29046.
31 B. Ghosh, B. Wu, H. K. Mulmudi, C. Guet, K. Weber, T. C. Sum, S. Mhaisalkar and N. Mathews, *ACS Appl. Mater. Interfaces*, 2018, **10**, 35000–35007.
32 N. K. Tailor, S. Mishra, T. Sharma, A. K. De and S. Satapathi, *J. Phys. Chem. C*, 2021, **125**, 9891–9898.
33 N. K. Tailor, P. Maity and S. Satapathi, *J. Phys. Chem. Lett.*, 2022, **13**, 5260–5266.
34 Y. Zhang, Y. Liu, Z. Xu, H. Ye, Z. Yang, J. You, M. Liu, Y. He, M. G. Kanatzidis and S. (Frank) Liu, *Nat. Commun.*, 2020, **11**, 2304.
35 S. Ghimire and C. Klinke, *Nanoscale*, 2021, **13**, 12394–12422.
36 S. Ghimire, K. Oldenburg, S. Bartling, R. Lesyuk and C. Klinke, *ACS Energy Lett.*, 2022, **7**, 975–983.

37 B. Yang, J. Chen, S. Yang, F. Hong, L. Sun, P. Han, T. Pullerits, W. Deng and K. Han, *Angew. Chem. Int. Ed.*, 2018, **57**, 5359–5363.
38 G. M. Paternò, N. Mishra, A. J. Barker, Z. Dang, G. Lanzani, L. Manna and A. Petrozza, *Adv. Funct. Mater.*, 2019, **29**, 1805299.
39 H. Zhang, Y. Xu, Q. Sun, J. Dong, Y. Lu, B. Zhang and W. Jie, *CrystEngComm*, 2018, **20**, 4935–4941.
40 A. Nilă, M. Baibarac, A. Matea, R. Mitran and I. Baltog, *Phys. Status Solidi B*, 2017, **254**, 1552805.
41 R. D. Nelson, K. Santra, Y. Wang, A. Hadi, J. W. Petrich and M. G. Panthani, *Chem. Commun.*, 2018, **54**, 3640–3643.
42 C. G. Bischak, E. M. Sanehira, J. T. Precht, J. M. Luther and N. S. Ginsberg, *Nano Lett.*, 2015, **15**, 4799–4807.
43 M. I. Dar, A. Hinderhofer, G. Jacopin, V. Belova, N. Arora, S. M. Zakeeruddin, F. Schreiber and M. Grätzel, *Adv. Funct. Mater.*, 2017, **27**, 1701433.
44 L. K. Jagadamma, P. R. Edwards, R. W. Martin, A. Ruseckas and I. D. W. Samuel, *ACS Appl. Energy Mater.*, 2021, **4**, 2707–2715.
45 D. Cortecchia, K. C. Lew, J.-K. So, A. Bruno and C. Soci, *Chem. Mater.*, 2017, **29**, 10088–10094.
46 X. Gonze, J.-M. Beuken, R. Caracas, F. Detraux, M. Fuchs, G.-M. Rignanese, L. Sindic, M. Verstraete, G. Zerah, F. Jollet, M. Torrent, A. Roy, M. Mikami, Ph. Ghosez, J.-Y. Raty and D. C. Allan, *Comput. Mater. Sci.*, 2002, **25**, 478–492.
47 X. Gonze, B. Amadon, P.-M. Anglade, J.-M. Beuken, F. Bottin, P. Boulanger, F. Bruneval, D. Caliste, R. Caracas, M. Côté, T. Deutsch, L. Genovese, Ph. Ghosez, M. Giantomassi, S. Goedecker, D. R. Hamann, P. Hermet, F. Jollet, G. Jomard, S. Leroux, M. Mancini, S. Mazevet, M. J. T. Oliveira, G. Onida, Y. Pouillon, T. Rangel, G.-M. Rignanese, D. Sangalli, R. Shaltaf, M. Torrent, M. J. Verstraete, G. Zerah and J. W. Zwanziger, *Comput. Phys. Commun.*, 2009, **180**, 2582–2615.
48 J. P. Perdew, K. Burke and M. Ernzerhof, *Phys. Rev. Lett.*, 1996, **77**, 3865–3868.
49 S. Rieger, B. J. Bohn, M. Döblinger, A. F. Richter, Y. Tong, K. Wang, P. Müller-Buschbaum, L. Polavarapu, L. Leppert, J. K. Stolarczyk and J. Feldmann, *Phys. Rev. B*, 2019, **100**, 201404.
50 D. Rossi, H. Wang, Y. Dong, T. Qiao, X. Qian and D. H. Son, *ACS Nano*, 2018, **12**, 12436–12443.
51 J. M. Frost, L. D. Whalley and A. Walsh, *ACS Energy Lett.*, 2017, **2**, 2647–2652.
52 W. P. D. Wong, J. Yin, B. Chaudhary, X. Y. Chin, D. Cortecchia, S.-Z. A. Lo, A. C. Grimsdale, O. F. Mohammed, G. Lanzani and C. Soci, *ACS Mater. Lett.*, 2020, **2**, 20–27.
53 C. Hartwigsen, S. Goedecker and J. Hutter, *Phys. Rev. B*, 1998, **58**, 3641–3662.

relax to the ground state in a few nanoseconds. Such delayed recombination of excitons can be utilized for catalysis.

**TOC Graphic**
Photoexcitation of $Cs_3Bi_2I_9$ nanosheets creates hot excitons which thermalize to the lattice in tens of picoseconds and finally

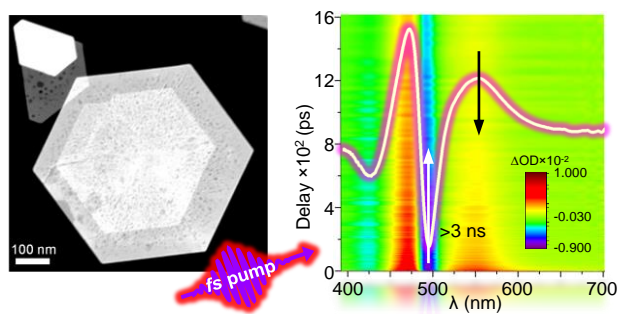